\documentclass{aa}
\usepackage{graphicx}
\usepackage{natbib,twoopt}
\usepackage[breaklinks=true]{hyperref} 
\bibpunct{(}{)}{;}{a}{}{,} 
\makeatletter
\newcommandtwoopt{\citeads}[3][][]{\href{http://adsabs.harvard.edu/abs/#3}%
{\def\hyper@linkstart##1##2{}%
\let\hyper@linkend\@empty\citealp[#1][#2]{#3}}}
\newcommandtwoopt{\citepads}[3][][]{\href{http://adsabs.harvard.edu/abs/#3}%
{\def\hyper@linkstart##1##2{}%
\let\hyper@linkend\@empty\citep[#1][#2]{#3}}}
\newcommandtwoopt{\citetads}[3][][]{\href{http://adsabs.harvard.edu/abs/#3}%
{\def\hyper@linkstart##1##2{}%
\let\hyper@linkend\@empty\citet[#1][#2]{#3}}}
\newcommandtwoopt{\citeyearads}[3][][]%
{\href{http://adsabs.harvard.edu/abs/#3}
{\def\hyper@linkstart##1##2{}%
\let\hyper@linkend\@empty\citeyear[#1][#2]{#3}}}
\makeatother
\usepackage[varg]{txfonts}
\usepackage{hyperref}
\begin{document} 

\title{Empirical instability strip for classical Cepheids: I.  The LMC galaxy}

   \subtitle{}

   \author{F. Espinoza-Arancibia\inst{1}
          \and
          B. Pilecki\inst{1}
          \and
          G. Pietrzy{\'n}ski\inst{1}
          \and
          R. Smolec\inst{1}
          \and
          P. Kervella\inst{2}
          }

   \institute{$^1$ Nicolaus Copernicus Astronomical Center, Polish Academy of Sciences, Bartycka 18, 00-716 Warsaw, Poland\\
    $^2$ LESIA (UMR 8109), Observatoire de Paris, PSL, CNRS, UPMC, Univ. Paris-Diderot, 5 place Jules Janssen, 92195 Meudon, France\\
              \email{fespinoza@camk.edu.pl}
             }

   \date{Received xxxx; accepted xxxx}

 
  \abstract
  %
   {The instability strip (IS) of classical Cepheids has been extensively studied theoretically. Comparison of the theoretical IS edges with those obtained empirically, using the most recent Cepheids catalogs available, can provide us with insights into the physical processes that determine the position of the IS boundaries.}
   {We investigate the empirical positions of the IS of the classical Cepheids in the Large Magellanic Cloud (LMC), considering any effect that increases its width, to obtain intrinsic edges that can be compared with theoretical models.}
   {We use data of classical fundamental-mode (F) and first-overtone (1O) LMC Cepheids from the OGLE-IV variable star catalog, together with a recent high-resolution reddening map from the literature. Our final sample includes 2058 F and 1387 1O Cepheids. We studied their position on the Hertzsprung-Russell diagram and determined the IS borders by tracing the edges of the color distribution along the strip.}
   {We obtain the blue and red edges of the IS in V- and I-photometric bands, in addition to $\log T_{\rm eff}$ and $\log L$. The results obtained show a break located at the Cepheids' period of about 3 days, which was not reported before. We compare our empirical borders with theoretical ones published in the literature obtaining a good agreement for specific parameter sets.}
   {The break in the IS borders is most likely explained by the depopulation of second and third crossing classical Cepheids in the faint part of the IS, since blue loops of evolutionary tracks in this mass range do not extend blueward enough to cross the IS at the LMC metallicity. Results from the comparison of our empirical borders with theoretical ones prove that our empirical IS is a useful tool for constraining theoretical models.}


   \maketitle

\section{Introduction}

During their evolution, intermediate- and high-mass stars can radially pulsate as Classical Cepheids (hereafter Cepheids) when they cross a defined region in the Hertzsprung-Russel diagram (HRD), called the instability strip (IS). Cepheids span a typical mass range between 3 and 13 ${\rm M}_{\sun}$ \citep[see, e.g.][]{Bono2000A,Anderson2016}, though Cepheids with masses higher than 13 ${\rm M}_{\sun}$ are also known \citep[][and references therein]{Musella2022}. With the exception of the most massive Cepheids, such stars first cross the IS during the H-shell burning phase, after they leave the main sequence (MS)\footnote{High mass stars ($\ga 9\mathrm{M}_{\sun}$) start helium-core burning on the way to the red giant branch (RGB) phase, during their first and only crossing of the IS \citep[see, e.g.][]{Anderson2016}}. Later, the star crosses the IS again during the helium-core burning phase, called the blue loop. The time scale of the first crossing is about 100 times shorter than that of subsequent crossings. Thus, it is assumed that most of the observed Cepheids are burning helium during the second or third crossing of the IS.

The location of the blue edge of the IS is related to the location of the partial ionization layers. These crucial zones, responsible for the excitation of radial pulsation by the so-called $\kappa$ and $\gamma$ mechanisms, could either be located too far out or not be present at all if the star is sufficiently hot \citep[see, e.g.,][and references therein]{catelan2015}. At the red end, convection produces a stabilizing effect which eventually quenches pulsation. Therefore, the exact location of the red edges depends on the complex interplay between pulsation and convection \cite[see][for a review and references]{Houdek2015}. Similarly, the IS has shown not to be really a ``strip" but rather to have a ``wedge" structure. The structure of the IS was reported initially using Galactic Cepheids by \cite{PelLub1978} and \cite{Fernie1990}, while in the Magellanic Clouds, it was noted by \cite{Martin1979}, and \cite{Caldwell1991}. A large number of theoretical studies have addressed the effects on the IS structure due to different physical properties. Among them, \cite{Bono2000B} showed how nonlinear theoretical models that consider the coupling between pulsation and convection produce an IS much less steep than linear models. \cite{Fiorentino2002}, and more recently \cite{Somma2022}, concluded that the IS location depends on both the metallicity and helium abundance, moving to lower effective temperatures as the metal content increases or the helium abundance decreases. \cite{Petroni2003} studied the effect of using different equations of state on the IS topology, finding a weak dependence. \cite{Fiorentino2007} explored the effects of varying the mixing-length parameter $l/H_{P}$, finding that the IS becomes narrower as $l/H_{P}$ increases, these results agree with the recent work done by \cite{Somma2022}. \cite{Anderson2016} conducted a study of the effects of rotation on Cepheid models, finding that the blue edge of the IS does not depend significantly on rotation, while the red edge is affected by metallicity and rotation. 

On the other hand, empirical studies have been conducted to study the properties of the IS. \cite{Turner2001} mapped the IS with a sample of 293 Cepheids in the Milky Way with reddening measurements, finding agreement with the results of \cite{PelLub1978}, but disagreement with those presented by \cite{Fernie1990}. \cite{Tammann2003} obtained period-color and period-luminosity (P-L) relations of a sample of 321 Galactic Cepheids, in addition to 314 and 486 Cepheids from the Large Magellanic Cloud (LMC) and the Small Magellanic Cloud (SMC), respectively. Using these relations, they obtained the IS of the Galaxy and the Magellanic Clouds, finding differences in the slopes of the ISs between the three galaxies. \cite{Sandage2004,Sandage2009} presented a detailed analysis of the samples of 593 and 460 Cepheids from the LMC and SMC, respectively. For both, they found breaks in the P-L relations, as well as in the IS edges, located at Cepheids' period of 10 days (for the LMC) and 2.5 days (for the SMC). Recently,  to find candidates for non-pulsating stars lying inside the IS, \cite{Narloch2019} determined a simple empirical instability strip for a sample of over 3200 LMC Cepheids. They used a step-detection technique and obtained almost parallel edges without a break. However, these authors did not take measures to clean the sample of outliers, and did not make any further analysis of the obtained IS.

Remarkably, the Optical Gravitational Lensing Experiment (OGLE) project presented an almost complete census of Cepheids in the Magellanic Clouds \citep{2017AcA....67..103S}. In addition, numerous reddening maps have been published in the literature \citep[see, e.g.][]{2016ApJ...832..176I,Gorski2020,2021ApJS..252...23S,2022MNRAS.511.1317C}. In this context, we aim to obtain an empirical intrinsic IS for the Cepheids in the LMC using the most recent Cepheids catalogs available. The comparison of theoretical IS edges with those obtained empirically can provide us with insights into the physical processes that determine the position of the IS boundaries. For this purpose, we use data of classical fundamental-mode (F) and first-overtone (1O) LMC Cepheids from the OGLE variable star catalog, together with recent high-resolution reddening maps from the literature. We study the position of the Cepheids on the color-magnitude diagram (CMD) and determine the IS borders by tracing their color distribution along the strip. In addition, we compare our empirical borders with theoretical ones presented in the literature.

The outline of this paper is as follows: Sect.~\ref{sec:sample} describes the sample selection procedure; Sect.~\ref{sec:ISborders} describes the method used to obtain the IS borders; Sect.~\ref{sec:discussion} presents a discussion of our results, including a comparison with LMC eclipsing binary stars and theoretical ISs published in the literature. Finally, Sect.~\ref{sec:conclusion} presents our conclusions.

\section{Sample selection}\label{sec:sample}
\begin{figure}
\centering
\includegraphics[width=\hsize]{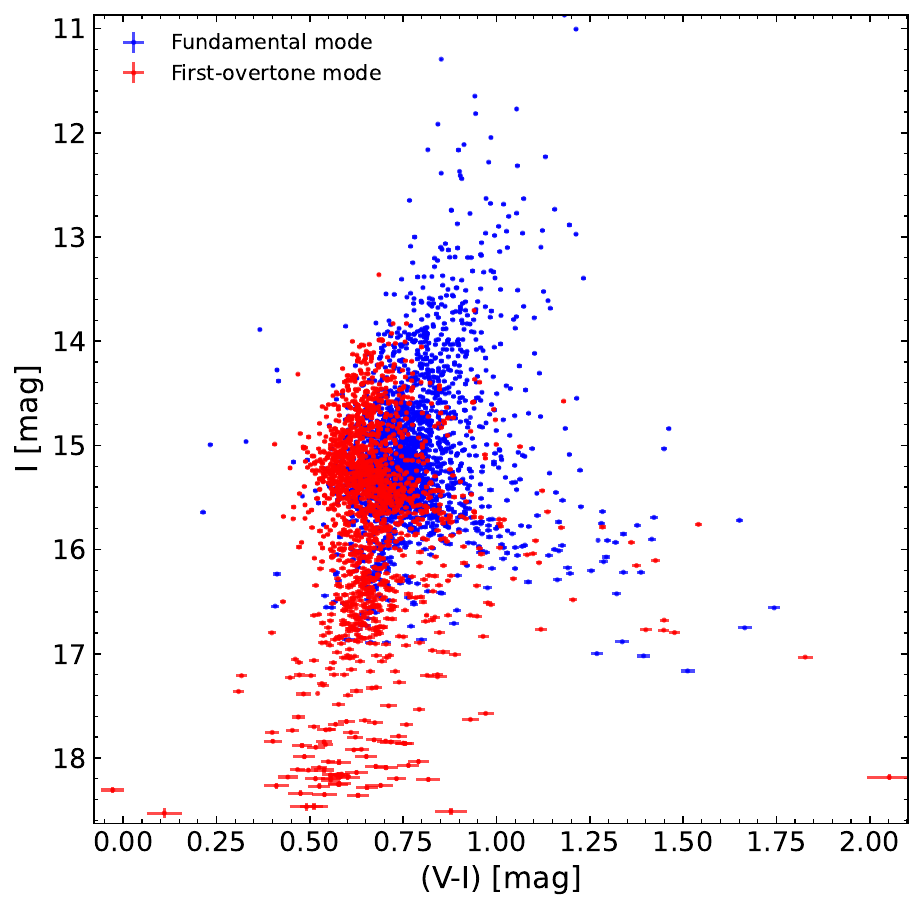}
\caption{Color-magnitude diagram of fundamental (blue) and first-overtone (red) LMC Cepheids from the OGLE-IV variable stars catalog.}
\label{fig1}
\end{figure}

We use data of F and 1O Cepheids in the LMC from the OGLE-IV variable stars catalog \citep{2017AcA....67..103S} and OGLE-III Shallow Survey \citep{Ulaczyk2013}. The distribution in the CMD of the full sample consisting of 2335 F Cepheids and 1682 1O Cepheids is shown in Fig.~\ref{fig1}.

The data is affected by reddening, produced by the interstellar extinction along the line of sight. To correct for this effect we use the high-resolution (1.7 arcmin $\times$ 1.7 arcmin) reddening map of \citet[]{2021ApJS..252...23S}, which is based on red clump stars from the Magellanic Clouds, selected from the OGLE-IV catalog. This map has shown good agreement with previous and other recent reddening maps \citep[e.g.][]{2016ApJ...832..176I,2022MNRAS.511.1317C}, and yields a small dispersion compared to other maps in the literature. It also covers both the central part and the outskirts of the LMC. These features are important when trying to recover the intrinsic IS, as discussed in Sect.~\ref{sec:ISborders}. We use the coefficients of relative extinction provided by \cite{Schlegel1998} to calculate the reddening correction.

\begin{figure}
\centering
\includegraphics[width=\hsize]{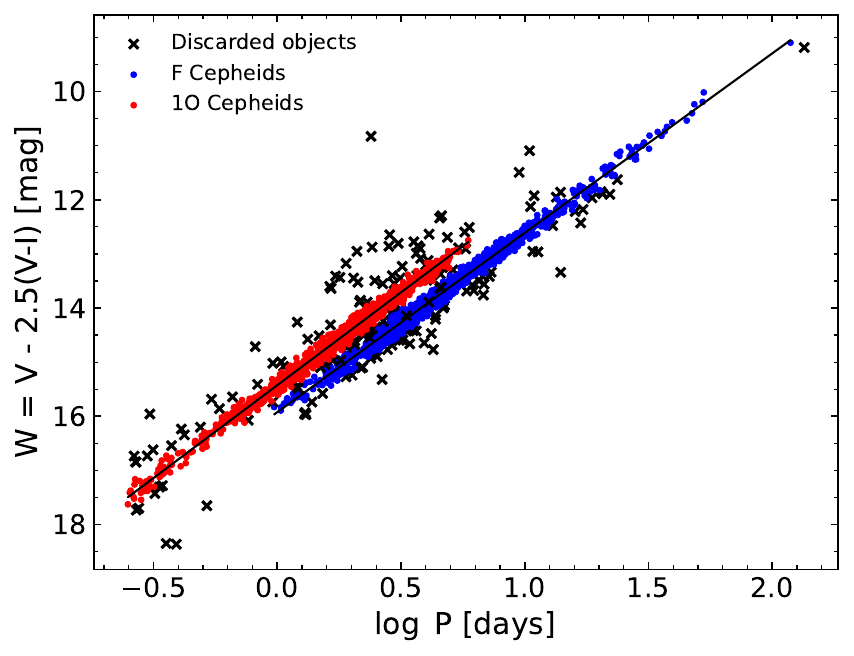}
\caption{Period-Wesenheit relations for fundamental (F, blue symbols) and first-overtone (1O, red symbols) LMC Cepheids. Discarded objects that deviate more than 3 sigma from the reddening-free P-L relation are marked with black crosses.}
\label{fig2}
\end{figure}

\begin{figure}
\centering
\includegraphics[width=\hsize]{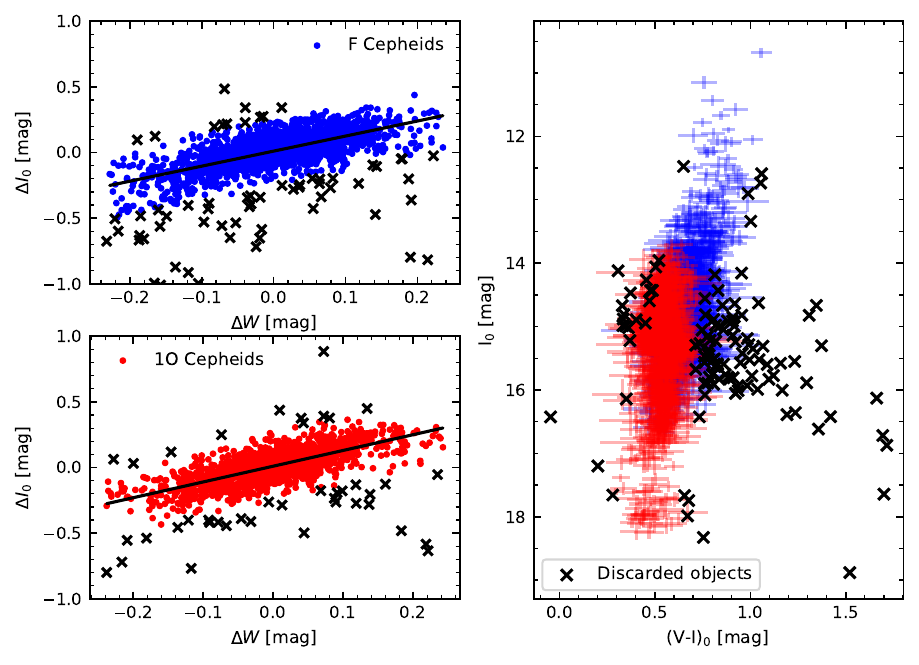} \caption{Diagram of magnitude residuals of the I-band P-L relation, $\Delta I_0$, against the corresponding residuals of the P-W relation, $\Delta W$, for F (upper left panel) and 1O (lower left panel) LMC Cepheids. Objects that deviate more than 3 sigma from this relation are marked with black crosses. These objects are also shown in a CMD of F (blue crosses) and 1O (red crosses) LMC Cepheids in the right panel.}
\label{fig3}
\end{figure}

In order to clean our sample, we discarded 164 objects deviating more than 3 sigma from the reddening-free period-luminosity relationship (also known as period-Wesenheit or P-W relation) as shown in Fig.~\ref{fig2}. A significant fraction of them are objects located significantly above the P-L relationship, which were identified recently to be mostly binary Cepheids with giant companions \citep{2021ApJ...910..118P, Pilecki2022b, Pilecki2022a}.
Moreover, we discarded 112 Cepheids that presented remarks in the OGLE-IV catalog, to avoid objects that have uncertain characterization. In addition, following the method described in \cite{2017ApJ...842...42M}, 113 Cepheids were not considered due to their high vertical deviation from the relation between the magnitude residuals of the I-band P-L relation and the corresponding residual of the P-W relation, as shown in Fig.~\ref{fig3}. These deviations are possibly due to errors in the individual adopted extinctions (as mentioned by Madore et al.). Since intrinsic color uncertainties are dominated by reddening uncertainties, as a final cleaning process, we did not consider Cepheids with reddening errors greater than the $95^{\rm th}$ percentile of the reddening error distribution. Consequently, 109 F Cepheids and 74 1O Cepheids were discarded.

\begin{figure}
\centering
\includegraphics[width=\hsize]{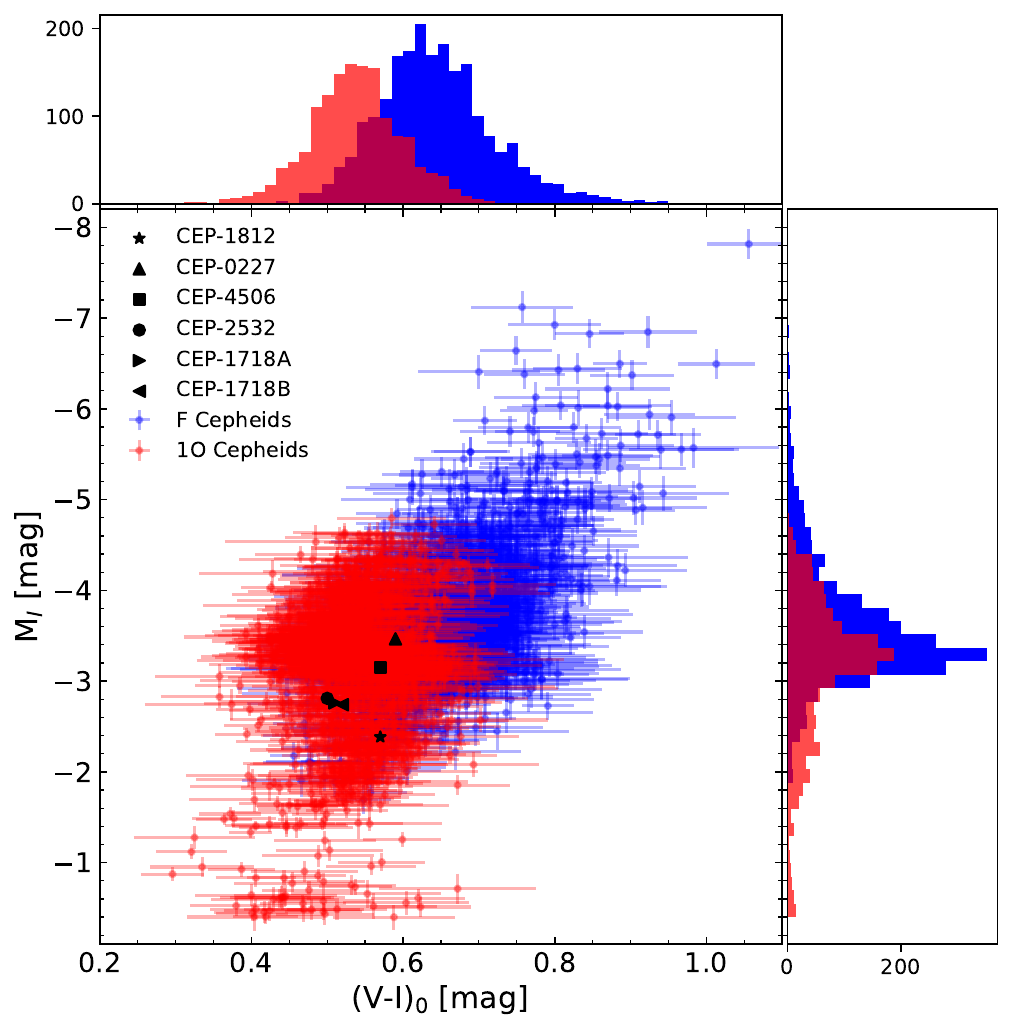}
\caption{CMD with I-band absolute magnitude for the final sample of F (blue) and 1O (red) LMC Cepheids. In different black symbols, data for well-documented Cepheids in eclipsing binaries are shown. Distributions of the intrinsic color (V-I)$_0$ and absolute magnitude $M_I$ are shown in the upper and right subpanels, respectively.}
\label{fig4}
\end{figure}

Individual distances to the stars, used to calculate their I-band absolute magnitudes, were computed using the method described in \cite{2016AcA....66..149J}, which takes into account the geometry of the LMC. We adopted their P-L relations for the full sample of F and 1O LMC Cepheids but only those without a break. As for the LMC distance, we used the recent accurate determination by \citet{Pietrzynski2019}. The distribution in the CMD of the final sample of 2058 F and 1387 1O Cepheids is shown in Fig.~\ref{fig4}. Additionally, the position in the CMD of well-documented F and 1O Cepheids in eclipsing binaries, presented in \cite{2018ApJ...862...43P}, was included.

\section{Instability Strip borders}\label{sec:ISborders}

The intrinsic width of the IS must be corrected for any effect that could change it, to be able to compare the strip with theoretical models. As we already removed from the sample all clear outliers (also those due to luminous companions) and objects with uncertain measurements, the major factor that is left is the effect of not perfectly known reddening. The non-zero uncertainties of the applied reddening values mean that the scatter of the Cepheid colors is higher and the IS width larger. The uncertainty of the photometry itself has a similar effect but to a much lesser extent. We performed the following steps to obtain an intrinsic IS that is free of this effect. We binned the complete sample of F and 1O Cepheids (the one shown in Fig. \ref{fig4}) with respect to absolute magnitude in the I-band $M_I$. The central bins contain 233 stars, while the faintest and brightest bins contain around 100 stars. Firstly, we determine the initials blue and red edges of the IS of each bin, by locating the $1^{\rm st}$ and $99^{\rm th}$ percentile of the intrinsic color distribution, namely $B_{\rm 1st}$ and $R_{\rm 99th}$, respectively. We then add random Gaussian errors to the intrinsic color of each Cepheid, making the distribution of color wider. Subsequently, we count the number of extra stars that after such procedure are located outside the initial edges, i.e., stars that are bluer or redder than $B_{\rm 1st}$ or $R_{\rm 99th}$, respectively. We repeated this process 10000 times and computed the median of the distribution of these numbers, namely $n_{\rm blue}$, $n_{\rm red}$. 

\begin{figure*}
\sidecaption
\includegraphics[width=0.7\linewidth]{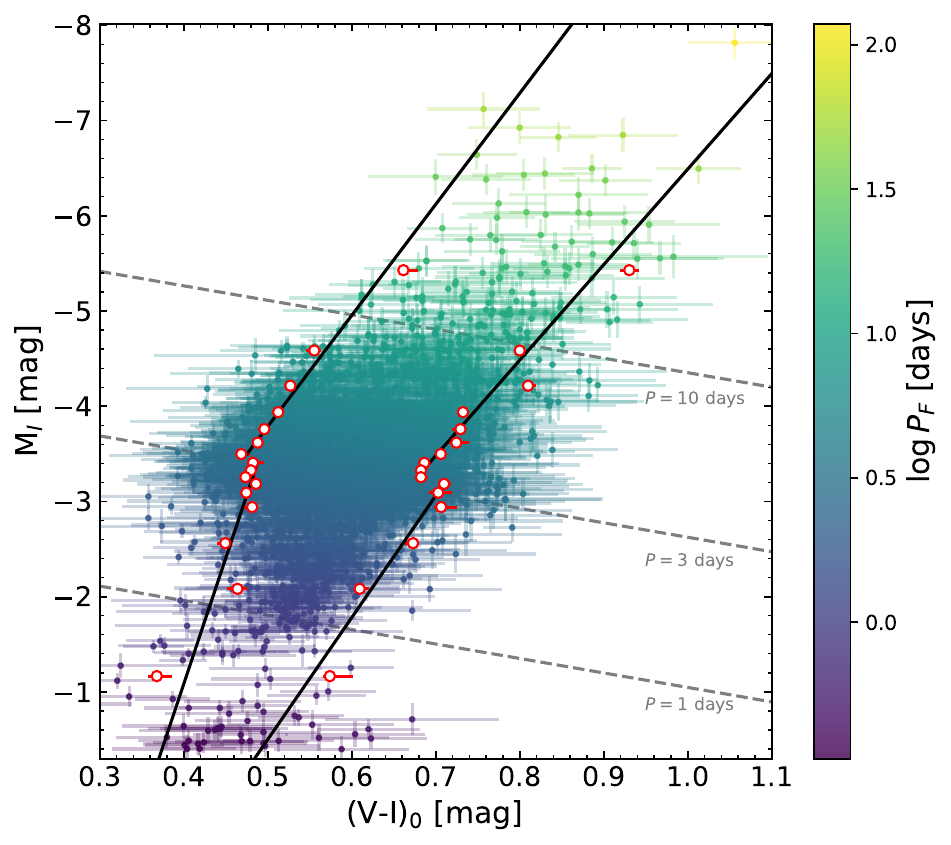}
\caption{CMD of F and 1O LMC Cepheids. The boundaries of our empirical IS are shown as red circles. Fits for the upper and lower part of the red and blue edges are shown as black solid lines. Periods for these stars are shown with a color gradient. For 1O Cepheids, periods were fundamentalized. Dashed lines of constant periods are overplotted. In most cases, error bars are smaller than the size of the points.}
\label{fig5}
\end{figure*}

\begin{figure*}
\resizebox{\hsize}{!}
{\includegraphics{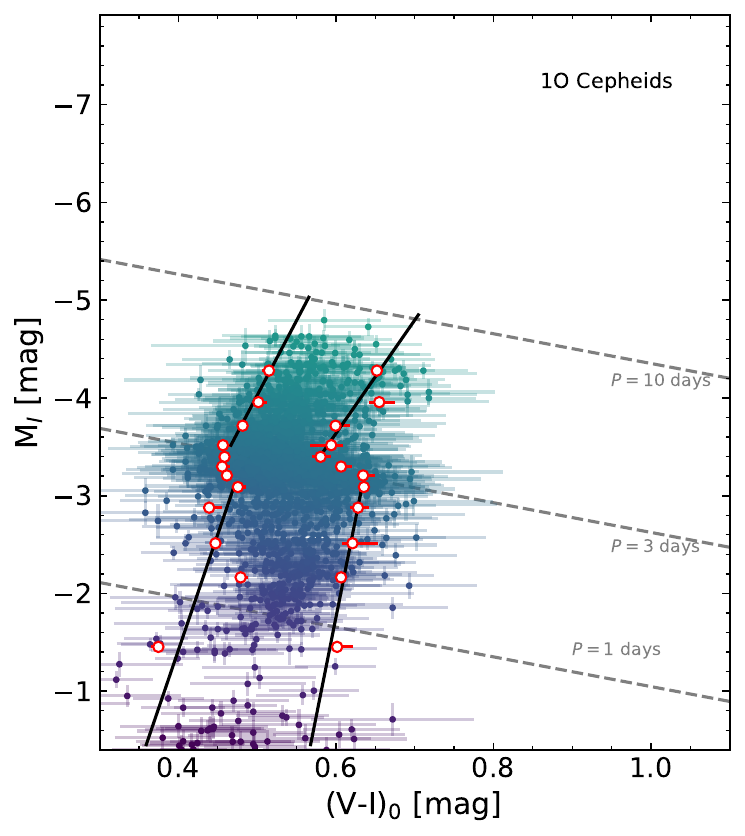}
\includegraphics{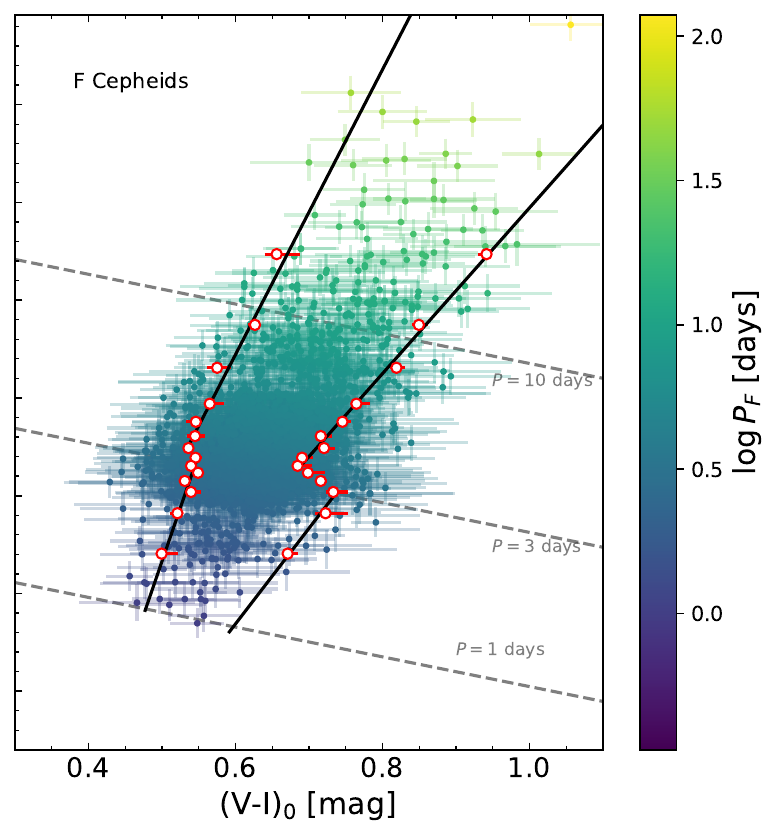}}
\caption{Same as Fig.~\ref{fig5}, for 1O (left panel) and F (right panel) independently.}
\label{fig6}
\end{figure*}

The final blue and red IS positions of each bin, namely, $B_{\rm final}$ and $R_{\rm final}$, were obtained by moving the initial edges ``inside" the IS by $n_{\rm blue}$ and $n_{\rm red}$ stars, respectively. In this way, we subtract the widening effect due to reddening uncertainties and obtain a better approximation to the intrinsic IS. Note that only such corrected IS can be compared directly with theoretical models that are not affected by reddening. The interior and exterior uncertainties in the intrinsic color of the IS were obtained by considering (instead of the median) the $16^{\rm th}$ and $84^{\rm th}$ percentile of the distribution of the number of extra stars outside the initial edges, $B_{\rm 1st}$ and $R_{\rm 99th}$. The positions in $M_I$ of each bin of the IS edges were considered as the median of the $M_I$ distribution of each bin. The computed IS boundaries, including F and 1O Cepheids, are shown in Fig.~\ref{fig5} as red circles. In this figure, the periods of the 1O Cepheids were fundamentalized using equation 1 of \cite{2021ApJ...910..118P}. To obtain the overplotted constant period lines for 1, 3, and 10 days we used the period-luminosity-color (PLC) relation computed using the previously calculated I-band absolute magnitude and intrinsic color, and the periods provided in the OGLE catalog. Along the 3-day constant period line, we notice a change of slope and discontinuity of the calculated edges between faint and bright Cepheids. A discussion of this phenomenon can be found in Sect. \ref{subsec:break}. Motivated by this detection, using the Python package \texttt{emcee} \citep{emcee}, we performed a linear fit with two parts for our empirical IS borders.
In addition, a simplified wedge-shaped IS version was also computed. Coefficients of both formulations of the ISs are listed in Table \ref{tab:table1}. We repeated the IS edge determination process for F and 1O Cepheids separately, using bins that contain 162 and 109 stars, respectively. The obtained IS edges, for F and 1O Cepheids individually, are shown in Fig.~\ref{fig6} as red circles. We computed linear fits for fainter and brighter Cepheids and a wedge-shaped IS as in Fig. \ref{fig5}. The corresponding coefficients of the ISs can be found in Table \ref{tab:table1}.

Additionally, we determined the IS edges in a different way, to check the accuracy of the method presented above. In the other approach after adding random errors to the intrinsic color of the stars, we computed the $1^{\rm st}$ and $99^{\rm th}$ percentile of the color distribution on each iteration, namely, $B^{i}_{\rm 1st}$ and $R^{i}_{\rm 99th}$ respectively. Afterward, we compared the mode of the distribution of $B^{i}_{\rm 1st}$ and $R^{i}_{\rm 99th}$ with the initial edges $B_{\rm 1st}$ and $R_{\rm 99th}$, calculating the respective differences ($\Delta B$ and $\Delta R$). The initial edges were then corrected by $\Delta B$ and $\Delta R$ towards the IS interior. Both presented approaches yielded consistent results within errors. Nevertheless, the first presented method (correction by number) produced a smaller scatter of edge points and with smaller uncertainties.

As the reddening is one of the most important factors in a proper positioning of the IS edges we did a check with two other reddening maps mentioned previously.
The reddening map of the central regions of the LMC and SMC presented by \cite{Gorski2020} was obtained using red clump stars selected from the data of OGLE-III, while similar maps of \cite{2022MNRAS.511.1317C} were derived from the SEDs of stellar sources using various large-scale surveys. We applied a reddening correction using their results and recomputed the IS edges of the sample including F and 1O Cepheids. The results were consistent with the ones obtained using the reddening map from \cite{2021ApJS..252...23S} (shown in Fig. \ref{fig5}). However, the blue edge obtained using the map of \cite{Gorski2020} yielded a mean shift towards bluer colors of 0.07 mag, while the red edge obtained using the map of \cite{2022MNRAS.511.1317C}  showed higher uncertainties.

\section{Discussion} \label{sec:discussion}

\subsection{Break in IS boundaries in (V-I) color}\label{subsec:break}
Independent from the used reddening maps, in all cases, a change in the slope and a discontinuity of the IS is observed at a period of 3 days, which makes this detection robust. We thus looked into possible explanations of this feature.
Partially, it can be produced by the change of the fraction of F to 1O Cepheids at this period ($P\sim3$ days, and M$_I\sim-3$) which can be seen in the histogram presented in Fig. \ref{fig4}. As the slopes for these modes differ (see Table~\ref{tab:table1}), such a change makes the IS less steep above the break. Nevertheless, this break is also present in the IS edges of the F and 1O Cepheid subsamples (most noticeable in the red edge) presented in Fig.~\ref{fig6} and thus cannot be attributed only to the fraction change.

Previously in the literature, breaks in the P-L relation of F and 1O Cepheids have been reported in Magellanic clouds. \cite{Bauer1999, Sharpee2002} and \cite{Sandage2009} found a slope change of the P-L relation for the SMC F Cepheids with periods shorter than 2 days. On the other hand, \cite{Sandage2004} and \cite{Ripepi2022} describe a break in the P-L relation for LMC Cepheids with periods of $P_{F}=10$ d and $P_{1O}=0.58$ d (the equivalent of $P_{F}=0.81$ d), respectively. \citet[][and references therein]{Bhardwaj2016} performed an analysis of possible non-linearity in the P-L relations of Cepheids in the Magellanic clouds, confirming the statistical significance of the aforementioned breaks.

\cite{Bauer1999} proposed several possible scenarios to explain the break of the P-L relation at 2 days for the SMC. One of these scenarios, also mentioned by \cite{Ripepi2022} in regard to the LMC, is the depopulation of second and third-crossing Cepheids in the faint part of the IS due to the lower extension of the blue loops as the mass and period of Cepheids decrease. These stars spend less time inside the IS, or they do not enter it at all. This implies that Cepheids fainter than the break are likely on their first crossing of the IS. Such a scenario should be noticeable in the CMD and can have an impact on the shape of the IS.

\begin{figure*}
\resizebox{\hsize}{!}
{\includegraphics{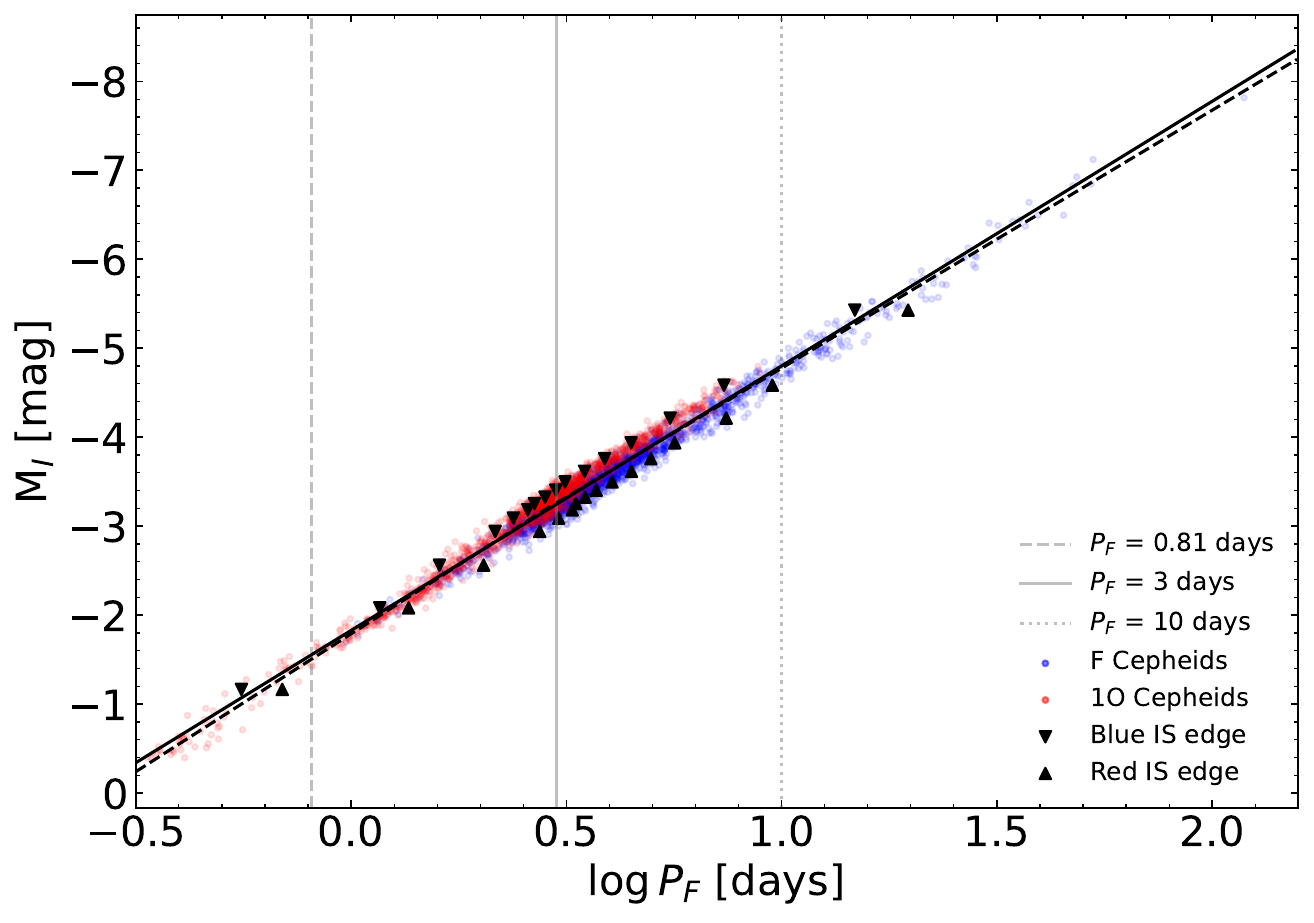}
\includegraphics{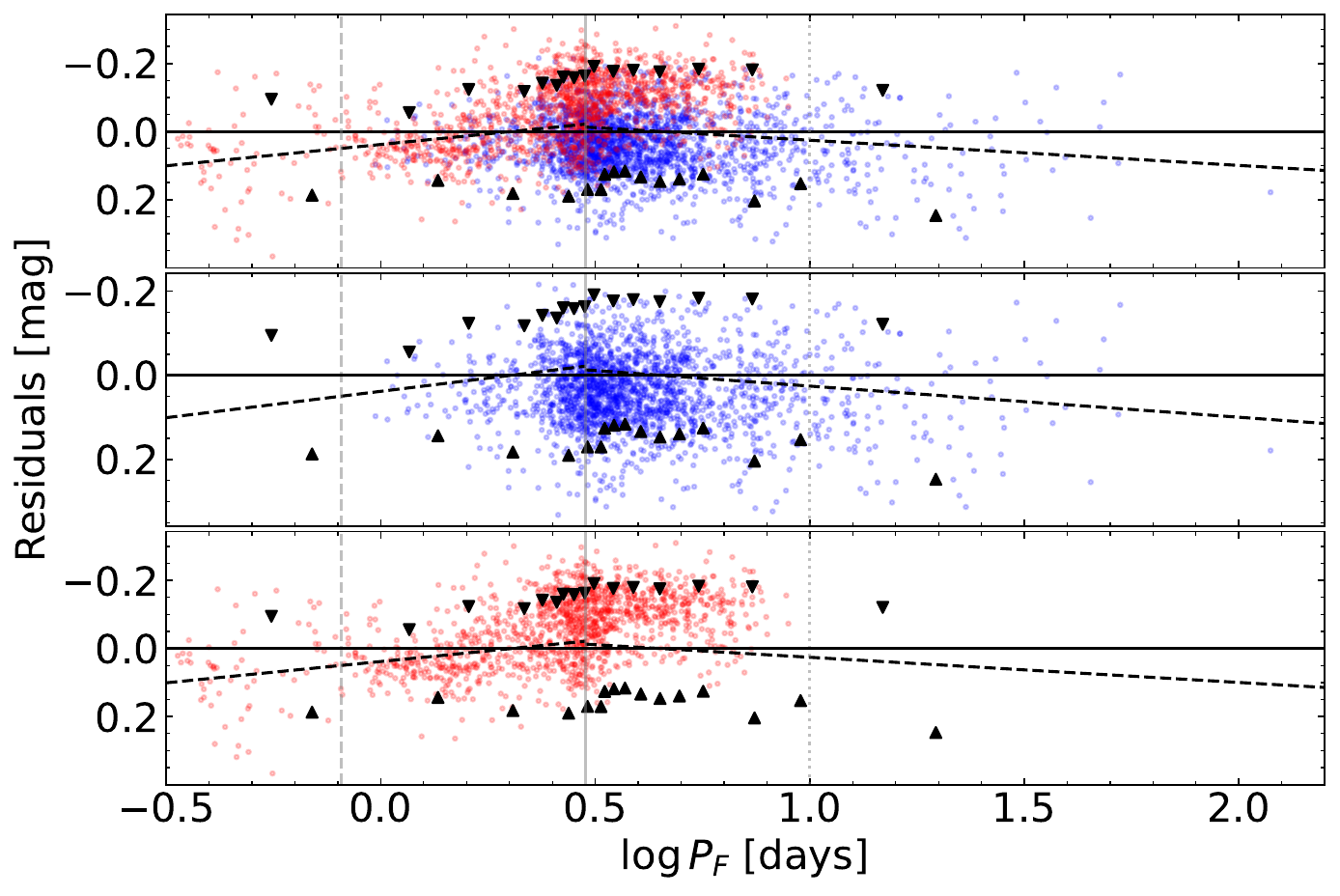}}
\caption{P-L relation of F (blue points) and 1O (red points) LMC Cepheids (left panel). The 1O Cepheid periods were fundamentalized. The boundaries of the IS are displayed as black points. At $P_F = 0.81, 3$ and $10$ days, a vertical gray line is traced. An average fit of the P-L relation considering the entire sample of stars and with a break at $P_F = 3$ days are shown as solid and dashed lines, respectively. The differences between the average fit of P-L with no break and the data are presented in the panels on the right, for the entire sample (upper), only F Cepheids (middle), and only 1O Cepheids (lower).}
\label{PLplot}
\end{figure*}

In Fig.~\ref{PLplot} we show an average P-L relation of our complete sample (considering F and 1O Cepheids), the periods of the 1O Cepheids were fundamentalized. In addition, a P-L relation with a break at $P=3$ days is shown. Using the PLC relation we converted our IS to the period-luminosity plane and marked its edges with black triangles. In the right panel, we show residuals from the subtraction of the fit of the P-L relation with no break from the data. For periods shorter than $P \sim 3$ days, there is a sudden decrease in the number of Cepheids. Stars with periods shorter than 2.5 days or longer than about 10 days tend to lay mostly below both P-L relations. The 10-day limit corresponds to the break reported in the literature, while below 0.81 days (a break from Ripepi et al.) we can only see another significant decrease in the number of Cepheids. Regarding the IS edges,  some change can be seen at these two period values but the statistics are too low to quantify it reliably. We have to admit that because of the joint analysis of the F and 1O Cepheids the observations presented in this paragraph may slightly depend on the fundamentalization method used but the results are consistent with what we can expect from the analysis of the color-magnitude diagram.

From a theoretical point of view, blue loops are very sensitive to metallicity and the adopted input physics, such as convective overshooting, and nuclear reactions \citep{2004Xua,2004Xub,Walmswell2015}. Several previous works in the literature have shown that the general trend for the blue loop extension is to decrease with stellar mass. In particular, for stars with $M<4~\mathrm{M}_{\sun}$ the loops do not enter the IS at all \citep[see, e.g.,][]{Anderson2016,Somma2022}. This supports the scenario of the depopulation of the faint part of the IS proposed by \cite{Bauer1999} and \cite{Ripepi2022}.

\begin{figure*}
\includegraphics[width=.47\textwidth]{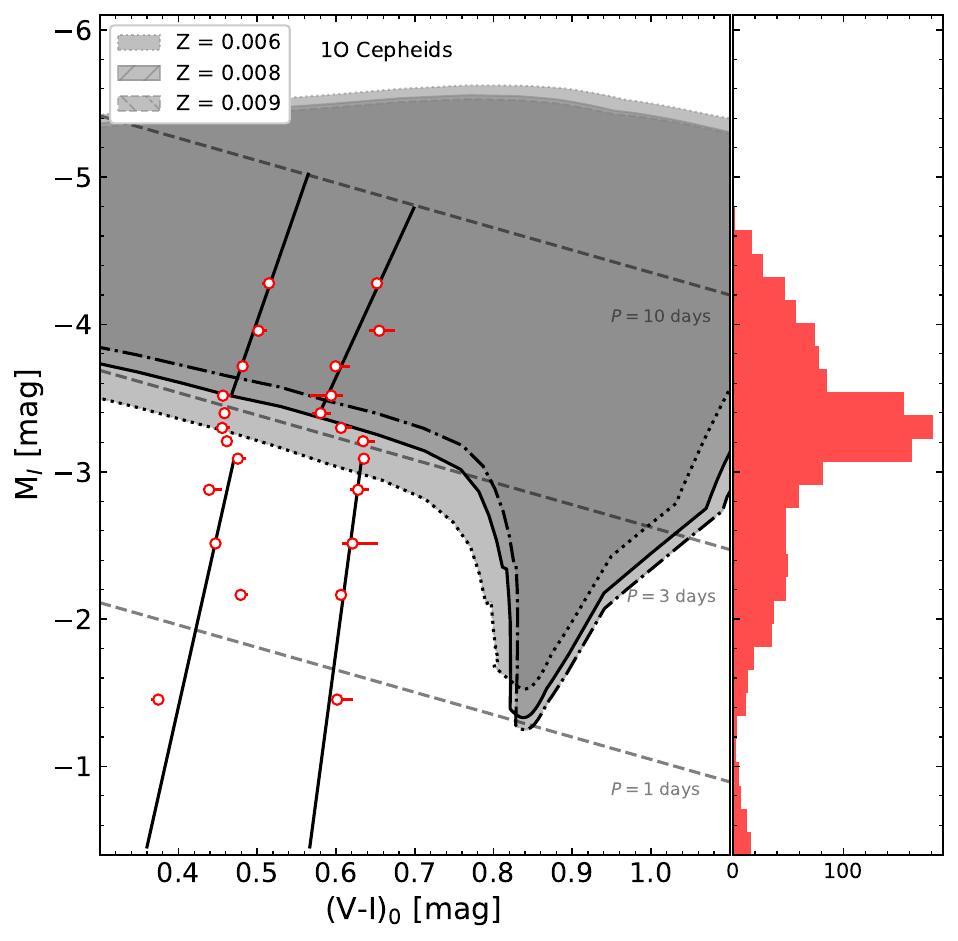}
\includegraphics[width=.47\textwidth]{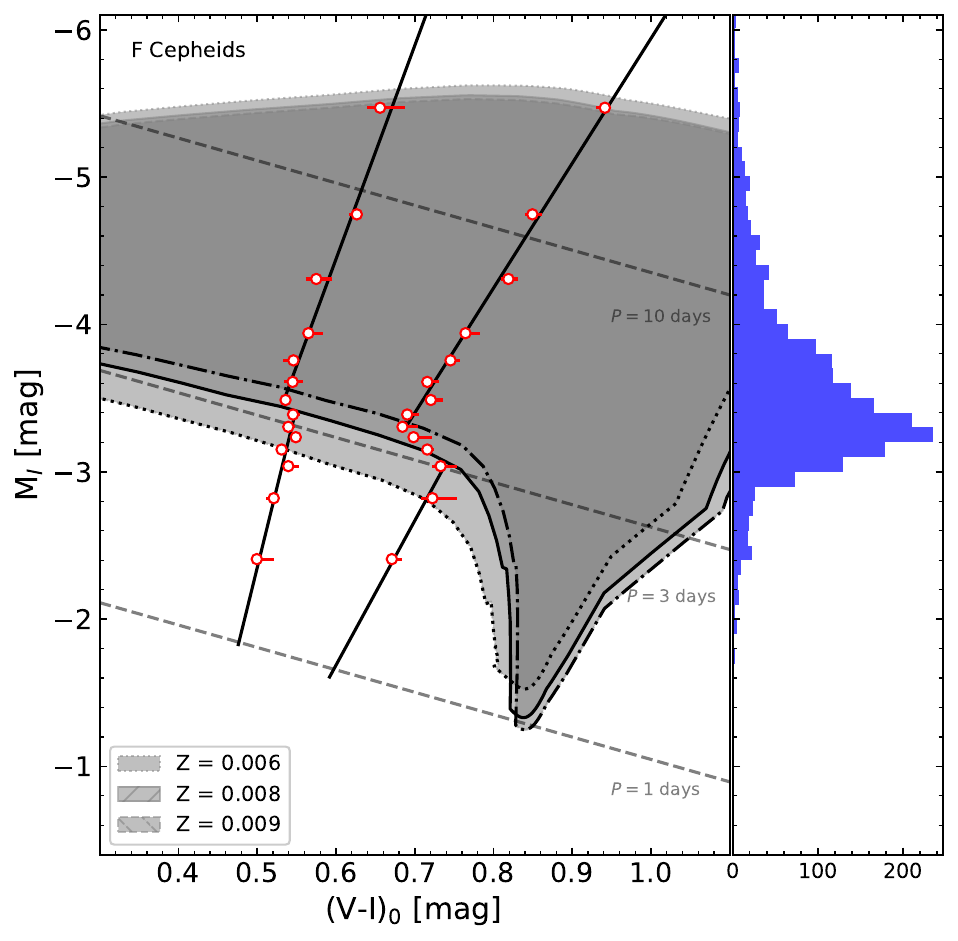}
\caption{CMD showing the empirical IS limits (red points), separately for 1O (left panel) and F-mode (right panel) LMC Cepheids. Fits for the blue and red edge, considering a break at $P\sim3$ days, are shown as solid lines. Shaded gray areas mark the blue loop extent (delimited by its bluest extreme and the tip of the RGB) for evolutionary tracks with masses from $3$ to $7$ M$_{\sun}$, and for $Z = 0.006$, $0.008$, and $0.009$. Note that the upper limits for the areas are defined by the evolutionary tracks for $7$ M$_{\sun}$ and would extend further up if tracks with higher masses are considered. The I-band absolute magnitude distributions of the 1O and F LMC Cepheids are shown on the right side of each panel.}
\label{fig10}
\end{figure*}

To further test theoretical models and the depopulation scenario,  we used the stellar evolution code Modules for Experiments in Stellar Astrophysics \citep[MESA;][]{Paxton2011,Paxton2013,Paxton2015,Paxton2018,Paxton2019,Jermyn2023} to calculate evolutionary tracks for non-rotating stars covering the mass range from 3 to 7~$\mathrm{M}_{\sun}$ in steps of 0.1~$\mathrm{M}_{\sun}$. We adopted  $Z=0.006$, $0.008$, and $0.009$ as representative metallicities for LMC stars \citep{Romaniello2022}. These tracks consider the solar mixture provided by \cite{Grevesse1998}. We use a solar-calibrated mixing length parameter of $\alpha_{\rm mlt} = 1.9$. For the convective boundaries, we use the predictive mixing scheme described in \cite{Paxton2018}, in addition to exponential core and envelope overshooting with the same parameter $f = 0.019$. The evolutionary tracks account for mass loss during the RGB, using the \cite{Reimers1975} prescription, with a scaling factor $\eta_R = 0.1$. The comparison between the evolutionary tracks and our empirical IS borders is shown in Fig.~\ref{fig10}, where we show the extent of blue loops (presented as shaded areas) for each adopted metallicity. These areas are delimited by the position of the tip of the RGB on the red side and the hottest extreme of the blue loop on the blue side. The upper limits are artificial and are defined by the evolutionary tracks for 7~$\mathrm{M}_{\sun}$-star with different metallicities. Evolutionary tracks with $M\lesssim 3.7$, $4.2$, and $4.3~\mathrm{M}_{\sun}$ for $Z = 0.006$, $0.007$, and $0.009$, respectively, show blue loops that do not pass through the red edge of our IS. These limits and the corresponding limits in the CMD are in agreement with the break in our IS, and in favor of the scenario of the depopulation of the second and third crossing Cepheids at the faintest end of the Cepheid IS. As a consequence, in these mass ranges, we expect an increasingly (toward shorter P) large contribution of Cepheids at the first crossing through the IS, where they evolve in a short time scale of the H-shell burning phase. Such first-crossing Cepheids spend less time within the IS compared to higher mass Cepheids, thus it is expected to find fewer stars in this part of the instability strip. On the other hand, evolutionary tracks for masses larger than the above-mentioned limits present blue loops that cross the red edge of the IS, therefore the fraction of Cepheids in the second and third crossings, evolving in the longer time scale of the He-core burning phase, increases considerably. Since these stars spend more time inside the strip, their total number is much larger than the stars in the lower part of the IS. From the coincidence with the detected break, we conclude that the evolutionary features mentioned above may be responsible for the shift of the faint IS edges to hotter temperatures and, partially, also for the change in the slopes (the latter being augmented by the change in the fraction of different modes).

\subsection{Break in IS boundaries in (V-K\texorpdfstring{$_s$}{s}) color}

\begin{figure}
\centering
\includegraphics[width=\hsize]{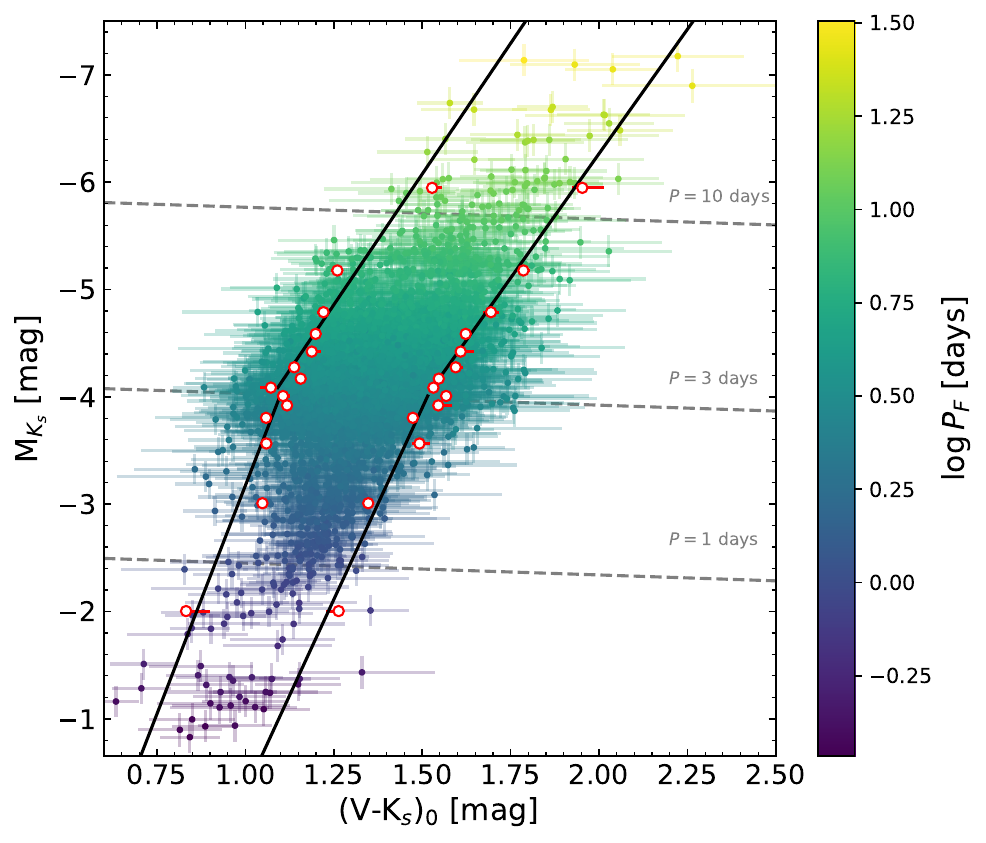}
\caption{Same as Fig.~\ref{fig5}, for V- and K$_s$-band.}
\label{VKCMD}
\end{figure}

Using the K$_s$-band data for Cepheids presented in \cite{Ripepi2022}, we explore whether the break in the positions of the IS for the V-I color index also appears for the V-K$_s$ index. We performed the same cleaning procedure mentioned in Sec.~\ref{sec:sample}, obtaining a total sample of 3168 LMC Cepheids. After applying the method described in Sect.~\ref{sec:ISborders}, we obtained empirical IS edges in the V and K$_s$-bands. Fig.~\ref{VKCMD} shows a CMD of the sample mentioned above, and determined edges of the empirical IS as individual points and as a linear fit to these points considering a break.
Due to the higher dependence of V-K$_s$ on temperature, the width of the IS in this color is larger than in the V-I but the uncertainties (mostly due to the reddening) are significantly higher, which eventually leads to a lower relative precision of the V-K$_s$ data. As the V-K$_s$ sample is also slightly less numerous we treat this analysis as auxiliary to that of the V-I data. Nevertheless, it is interesting to compare the results.

Similarly to the case with the V and I-bands, the red border of the IS shows a clear change in the slope at $M_{K_s}\sim-4$, a value that corresponds to $P\sim3$ days, approximately. Interestingly, contrary to the analysis with the V-I color, with the V-K$_s$ index at the break there is no significant shift of the red edge towards redder colors for shorter-period Cepheids.
Analyzing the IS borders in the V and I bands, we identified two main factors responsible for the break: the change in the fraction of F to 1O Cepheids, and the depopulation of 2nd and 3rd crossing Cepheids in the lower part of the IS.
Surely, the first factor applies to the V-K$_s$ color as well, as the edges for F Cepheids (analyzed separately) are less steep than for 1O Cepheids also in these bands. The effect of the evolutionary factor is harder to evaluate. The use of the I-band apparently enhances the difference between the Cepheids with periods shorter and longer than 3 days. However, if the I-band is for some reason more sensitive to the evolutionary stage (Hertzsprung gap vs. blue loop) than the K$_s$-band, would need another study. One has to also note that both the wider IS and the higher uncertainties in V-K may, to some extent, hinder the detection of discontinuity.

\begin{table*}
\centering
\begin{tabular}{|c|c|c|c|c|c|c|c|c|c}
\hline\hline
        $P$ [d]&$\alpha_{\rm blue}$&$\beta_{\rm blue}$&$ \sigma_{\alpha,{\rm blue}}$&$\sigma_{\beta,{\rm blue}}$&$\alpha_{\rm red}$&$\beta_{\rm red}$&$\sigma_{\alpha,{\rm red}}$&$\sigma_{\beta,{\rm red}}$\\\hline
        \multicolumn{9}{|c|}{F and 1O Cepheids} \\\hline
        $< 3$ d &-0.038 & 0.491 & 0.001 & 0.001 & -0.078 & 0.734 & 0.003 & 0.004\\
        $> 3$ d &-0.086 & 0.475 & 0.002 & 0.001 & -0.100 & 0.702 & 0.001 & 0.001\\
        all &-0.059 & 0.490 & 0.0004 & 0.0002 & -0.076 & 0.71 & 0.001 & 0.0004\\\hline
        \multicolumn{9}{|c|}{F Cepheids} \\\hline
        $< 3$ d &-0.046 & 0.553 & 0.007 & 0.003 & -0.102 & 0.785 & 0.022 & 0.020\\
        $> 3$ d &-0.069 & 0.535 & 0.005 & 0.004 & -0.118 & 0.712 & 0.003 & 0.002\\
        all &-0.043 & 0.553 & 0.002 & 0.001 & -0.089 & 0.727 & 0.001 & 0.001\\\hline
        \multicolumn{9}{|c|}{1O Cepheids} \\\hline
        $< 3$ d & -0.042 & 0.488 & 0.002 & 0.003 & -0.025 & 0.643 & 0.003 & 0.004\\
        $> 3$ d & -0.065 & 0.466 & 0.006 & 0.004 & -0.086 & 0.588 & 0.009 & 0.004\\
        all & -0.038 & 0.482 & 0.001 & 0.001 & -0.014 & 0.625 & 0.001 & 0.001\\\hline\hline
\end{tabular}
\caption{Coefficients of the red and blue edges of the IS considering a break at $P = 3$ d, assuming $(V-I)_{0}=\alpha (M_{I} + 3.5)+\beta$, for F and 1O mode together and separately. In addition, we include the coefficients for a fit without a break.}
\label{tab:table1}
\end{table*}

\subsection{Comparison with LMC eclipsing binary stars}

\begin{figure}
\centering
\includegraphics[width=\hsize]{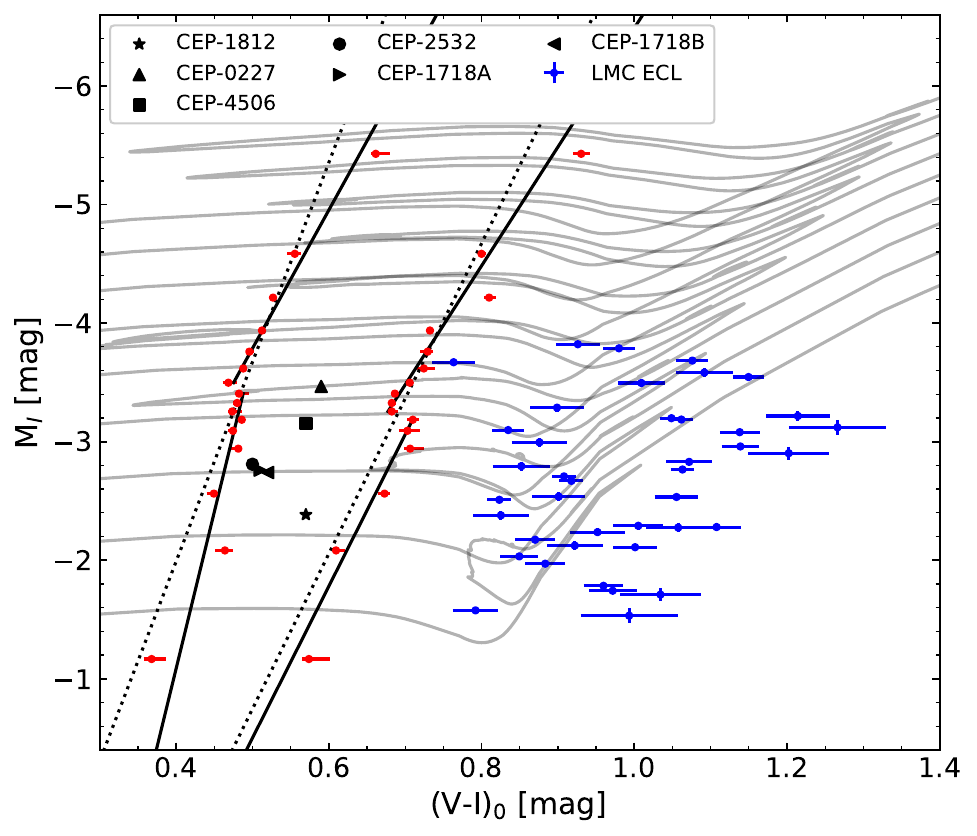}
\caption{CMD comparing our empirical IS (red points) with LMC eclipsing binary stars (black points). Data for well-documented Cepheids in eclipsing
binaries are displayed as black symbols. Fits for the blue and red edges of the IS are shown as solid lines (considering a break) and dotted lines (without a break). In addition, evolutionary tracks of $3$ to $7$ M$_{\sun}$ in steps of $0.5$ M$_\sun$ are presented as gray lines.}
\label{fig11}
\end{figure}

\cite{2018ApJ...860....1G} determined precise physical parameters of 20 eclipsing detached, double-lined LMC binary systems containing giant stars. Using the magnitudes in the I-band, reddening, and third light provided in that work, we calculate the absolute magnitude in the I-band and the intrinsic color $(V-I)_0$ for each component of the 20 systems. A CMD of the components of these systems, together with the two formulations (with and without a break) for the IS calculated in this work, is presented in Fig.~\ref{fig11}. As expected, these stars lie outside our computed IS, since the components of these binary systems are G- and early K-type giant stars. Remarkably, one star of the system OGLE LMC-ECL-13360 is slightly outside the fitted lines of the red edge for our two versions of the IS within errors. \cite{2018ApJ...860....1G} showed that it is possible to fit both stars of this system with an isochrone without rotation. This fit indicates that the star closer to our IS is in the helium-core burning phase, entering the IS, and would eventually start pulsating as a Cepheid.

\subsection{Comparison with theoretical ISs}

Previous works in the literature have obtained theoretical IS edges and have studied the effects of, for example, rotation, metallicity, and overshooting. By comparing these theoretical results with our empirical IS, it is possible to constrain some of the physical effects studied.

Each point of our IS for F Cepheids in the CMD plane was converted to the $\log{T_{\rm eff}}-\log{L}$ plane using the color-temperature and bolometric correction calibration presented by \cite{Worthey2011}. We averaged the effective temperature and luminosity calculated for each point, considering a surface gravity of $\log g = 1, 1.5, 2$ and $2.5$. Then the points were fitted in the same way as for the IS in the color-magnitude diagram. The coefficients of the empirical instability strip in the $\log{T_{\rm eff}}-\log{L}$ plane are presented in Table~\ref{tab:table2}. Since our empirical edges are composed of two parts, the conclusions may be different for the lower and upper parts of the IS. Additionally, our empirical results are based on a sample of Cepheids which have a non-zero spread in metallicity, which contributes to the uncertainty when comparing with theoretical models with a fixed metallicity.

\begin{figure*}
\includegraphics[width=.47\textwidth]{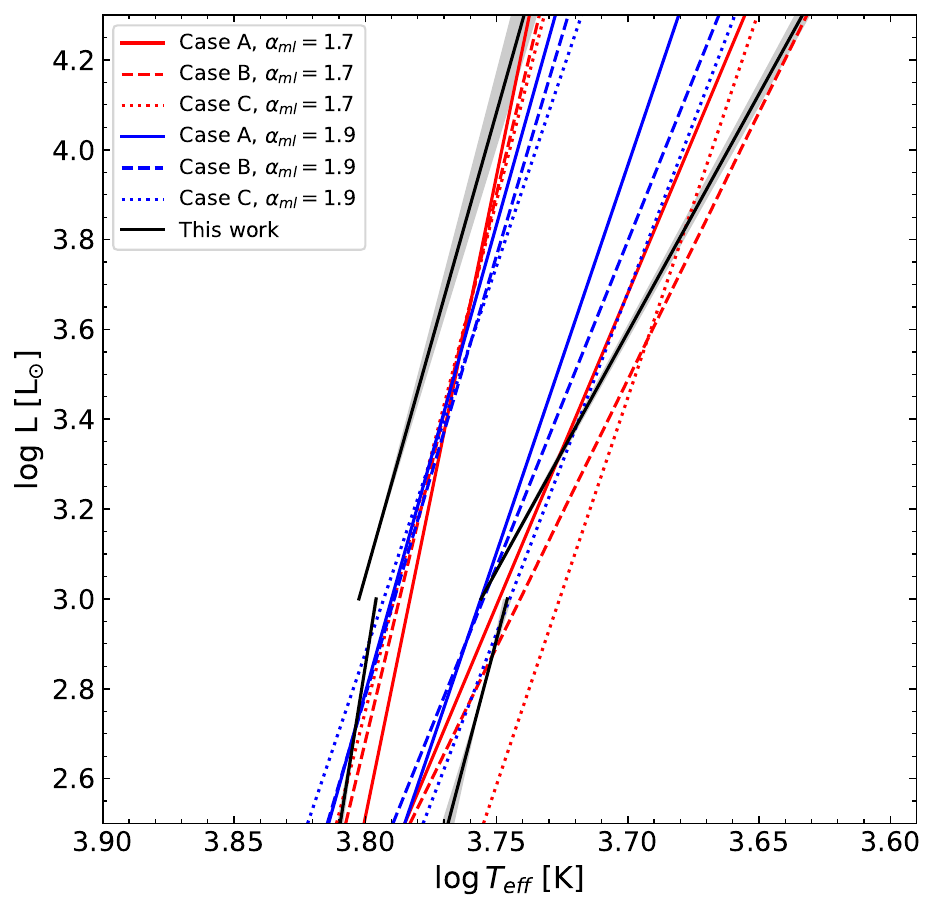}
\includegraphics[width=.47\textwidth]{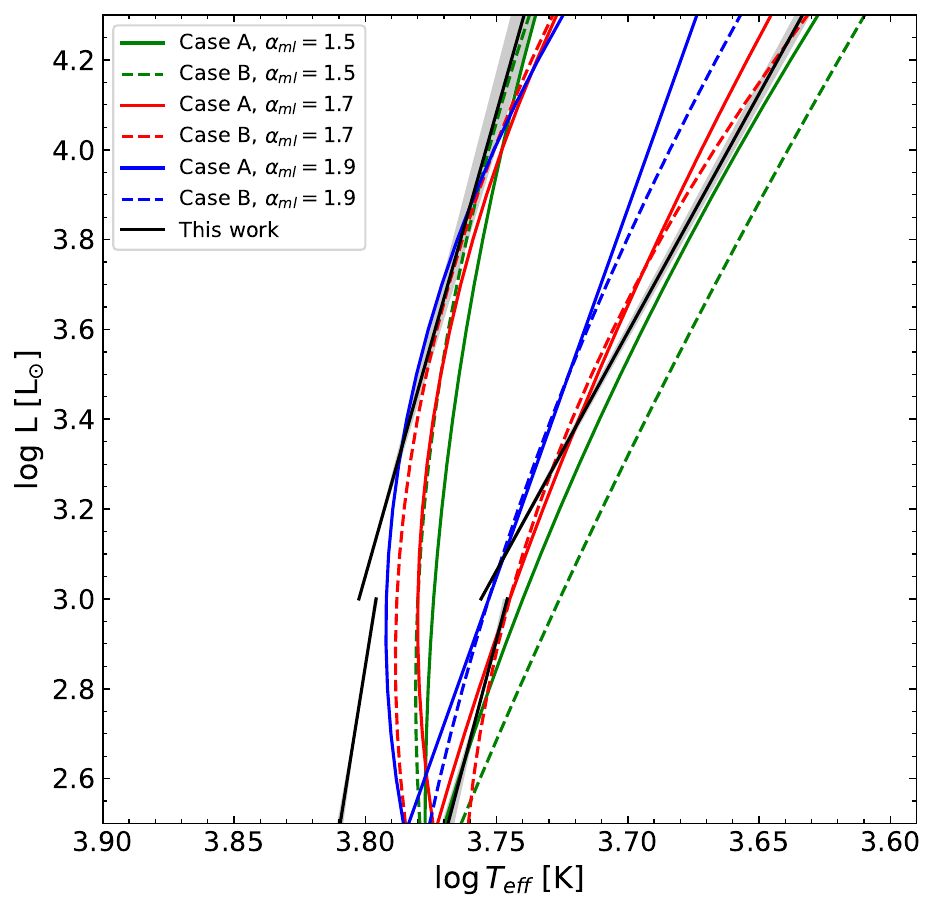}
\caption{Comparison of theoretical ISs considering linear (left panel) and quadratic (right panel) relations for $Z = 0.008$ presented in \cite{Somma2022} (red, blue and green) and our empirical IS (black) in an HRD. Different line styles represent different cases of the luminosity adjustment (increase) applied by the authors to their canonical model. Different colors represent different efficiencies of superadiabatic convection $\alpha_{ml}$. Shaded areas represent 1-sigma errors on the edges.}
\label{fig:DeSomma}
\end{figure*}

In Fig.~\ref{fig:DeSomma}, we show the theoretical ISs for F Cepheids obtained by \cite{Somma2022}. Their models are based on Bag of Stellar Tracks and Isochrones \citep[BaSTI;][]{Hidalgo2018} evolutionary tracks that take into account instantaneous core overshooting with a free parameter equal to 0.2, mass loss with the Reimers formula considering $\eta = 0.3$, and a metallicity $Z = 0.008$. As far as their nonlinear pulsation calculations go, \citet{Somma2022} approximate the effects of overshooting by considering an increase in the luminosity, over their canonical models (case A), of $\Delta(\log L/L_{\sun}) = 0.2$~dex (case B), and $\Delta(\log L/L_{\sun}) = 0.4$~dex (case C), considering in addition 3 values for the superadiabatic convection efficiency, $\alpha_{ml}=1.5$, $\alpha_{ml}=1.7$ and $\alpha_{ml}=1.9$. In the figure we show the linear and quadratic relations, that describe the theoretical IS boundaries, provided by the authors. In the left panel, for clarity, we exclude their results for $\alpha_{ml}=1.5$ given a weak agreement with our results. Globally, our empirical edges are most consistent with their models for case B and $\alpha_{\rm ml}=1.7$. However, there are some noticeable differences -- their linear relations are systematically cooler than our upper blue edge, and their quadratic relations are systematically cooler than our lower blue edge. Moreover, even the edges that coincide better with the empirical ones have slightly different slopes.

\begin{figure}
\centering
\includegraphics[width=\hsize]{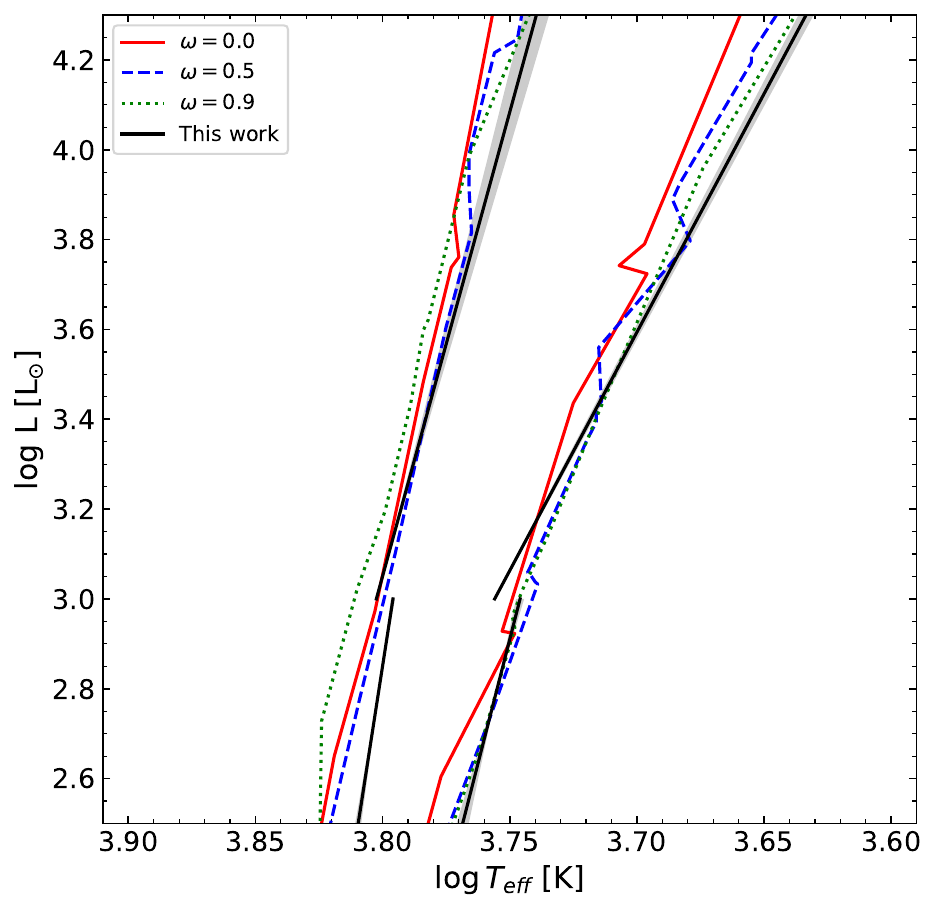}
   \caption{Comparison of theoretical ISs for $Z=0.006$ presented in \cite{Anderson2016} (red, blue, and green) and our empirical IS (black) in an HRD. Different line styles represent different rotation rates $\omega$, expressed as a fraction of critical velocity on ZAMS. Shaded areas represent 1-sigma errors on the edges.}
      \label{fig:Anderson}
\end{figure}

The theoretical ISs for F Cepheids of \cite{Anderson2016} are shown in Fig.~\ref{fig:Anderson}. They use the Geneva code of stellar evolution \citep{Eggenberger2008}, varying metallicity, and rate of rotation. We compare our results with their ISs for $Z=0.006$, and three rotation rates $\omega = 0.0$, $0.5$, and $0.9$. We reproduced their IS boundaries from Table A.2, where the effective temperature and luminosity of model stars entering or exiting the IS are given for each mass. For comparison with our empirical results, we used their results for the three crossings of the IS (if available) for all the provided masses. Their theoretical edges agree remarkably well with our empirical ones, both in regard to the zero points and the slopes, only with our lower blue edge being systematically cooler. One can note, that their red edges present several shifts to higher temperatures. This is related to the dependency of the red IS boundary of the crossing number. After the first crossing, structural changes in the star affect the way convection operates. In addition, an increase in helium content in the envelope occurs as a consequence of the first dredge-up. Both effects impact the position of the cool IS edge \citep{Anderson2016}. Globally their models with high rotation rates describe best our determined edges but the lower part seems to favor those with moderate rotation rates. 

\begin{figure}
\centering
\includegraphics[width=\hsize]{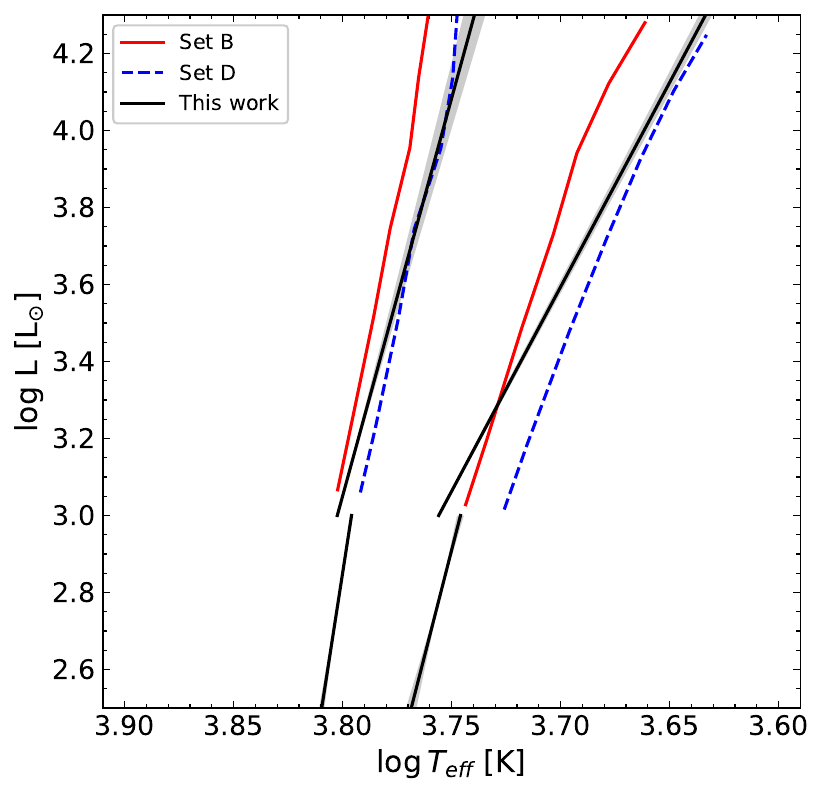}
   \caption{Comparison of theoretical ISs for $Z=0.008$ presented in \cite{Paxton2019} (red and blue) and our empirical IS (black) in an HRD. Different line styles represent different sets of convective parameters of RSP. Shaded areas represent 1-sigma errors on the edges.}
      \label{fig:MESA}
\end{figure}

\cite{Paxton2019} introduce Radial Stellar Pulsation \texttt{RSP}, a functionality in the MESA code to model high-amplitude, self-excited, non-linear pulsations that the star develops when it crosses the IS. The \texttt{RSP} model depends on equations describing time-dependent convection from \cite{Kuhfuss1986} and implemented by \cite{Smolec2008}, which also depend on free parameters. Pulsation periods depend weakly on these parameters. However, period growth rates and light curves are sensitive to the choice of these convective variables. Different sets for these parameters are shown in Table 4 of \cite{Paxton2019}. In Fig.~\ref{fig:MESA}, the theoretical ISs borders presented in \cite[][priv. communication]{Paxton2019} considering set B and D, for $Z=0.008$ are shown. Their ISs were computed using second and third-crossing Cepheid models. Set B of convective parameters adds radiative cooling effects to the convective model, while parameter set D simultaneously includes the effects of turbulent pressure and turbulent kinetic energy flux in addition to radiative cooling. As mentioned in \cite{Paxton2019}, the IS edges produced by sets A and C overlap those for convective sets B and D, respectively. We can see that our blue edge is in closer agreement with that of set D, while our red edge lies somewhere in between sets B and D. The red edge is actually described better by set D for luminous long-period Cepheids, and by set B for short-period Cepheids, also if we extrapolate their red edge down to fainter Cepheids.


\begin{table*}
\centering
\begin{tabular}{|c|c|c|c|c|c|c|c|c|c}
\hline\hline
        $P$ [d]&$\alpha_{\rm blue}$&$\beta_{\rm blue}$&$ \sigma_{\alpha,{\rm blue}}$&$\sigma_{\beta,{\rm blue}}$&$\alpha_{\rm red}$&$\beta_{\rm red}$&$\sigma_{\alpha,{\rm red}}$&$\sigma_{\beta,{\rm red}}$\\\hline
        \multicolumn{9}{|c|}{F and 1O Cepheids} \\\hline
        $< 3$ d &-0.021 & 3.805 & 0.001 & 0.0004 & -0.057 & 3.732 & 0.002 & 0.002\\
        $> 3$ d &-0.058 & 3.800 & 0.001 & 0.0001 & -0.074 & 3.737 & 0.001 & 0.0002\\
        all & -0.035 & 3.802 & 0.0002 & 0.0001 & -0.055 & 3.740 & 0.0004 & 0.0001\\\hline
        \multicolumn{9}{|c|}{F Cepheids} \\\hline
        $< 3$ d &-0.027 & 3.788 & 0.004 & 0.002 & -0.045 & 3.732 & 0.008 & 0.004\\
        $> 3$ d &-0.048 & 3.788 & 0.004 & 0.001 & -0.094 & 3.728 & 0.002 & 0.001\\
        all &-0.032 & 3.787 & 0.001 & 0.0004 & -0.072 & 3.729 & 0.0004 & 0.0002\\\hline
        \multicolumn{9}{|c|}{1O Cepheids} \\\hline
        $< 3$ d & -0.030 & 3.801 & 0.001 & 0.001 & -0.017 & 3.768 & 0.003 & 0.002\\
        $> 3$ d & -0.038 & 3.807 & 0.004 & 0.0004 & -0.062 & 3.774 & 0.007 & 0.001\\
        all & -0.023 & 3.807 & 0.001 & 0.0002 & -0.009 & 3.776 & 0.001 & 0.0002\\\hline\hline
\end{tabular}
\caption{Coefficients of the red and blue edges of the IS considering a break at $P = 3$ d, assuming $\log{T_{\rm eff}}=\alpha(\log {L} -3.3)+\beta$, for F and 1O together and separately. In addition, we include the coefficients without break.}
\label{tab:table2}
\end{table*}

\section{Conclusions}\label{sec:conclusion}

We used a sample of LMC Cepheids from the OGLE-IV catalog to compute the empirical, intrinsic instability strip. We determined the IS edges for F and 1O Cepheids, together and separately, using I and V-bands, and additionally using V and K$_s$-bands. In all cases, a break in the IS was observed at $\log{P}\sim 3$ days, which is most noticeable in the red edges and for separate 1O and F samples.

Using a set of evolutionary tracks computed with the stellar evolution code MESA, we investigated the scenario in which the origin of this break is explained by depopulation in the faint part of the IS. Such depopulation would be caused by too low an extension of blue loops for lower-mass stars, which would not evolve blueward enough to enter the instability strip. Our results confirmed the conclusions from other studies in the literature on that subject \citep{Bauer1999,Ripepi2022}.

Consequently, Cepheids with masses below $\sim 4 M_\sun$ would be found mostly during the first crossing of the IS, while above this limit there is a much higher contribution of Cepheids at the second and third crossing, for which the time scales are much longer. This change from the domination of the first to the domination of subsequent crossings coincides with the observed break in the IS edges and in the P-L relation. At the same time at different crossings, we may expect slightly different pulsational properties as the Cepheids are found at different evolutionary stages, i.e. the subgiant and blue loop phases of evolution (see Fig.~\ref{fig10}.). 
These two populations of Cepheids with different properties dominating at a given mass (or period) range, may then affect the position of the IS edges.

We compared our results with theoretical instability strips published in the literature. Our empirical edges are best described by the models of \citet{Anderson2016} with high rotation rates, for periods above the break, and by models with moderate rotation rates below the break. However, regarding the red edge only, models with $\omega=0.9$ describe well the whole range of pulsation periods.

Regarding the instability strips that result from the models of \citet{Somma2022} and \citet{Paxton2019}, some fine-tuning of parameters would be needed to obtain consistent results. At the moment, no single set of their parameters produces an IS that would correspond well with our empirical edges -- they either fall between the different sets or have different slopes. 

Such fine-tuning is beyond the scope of this paper but our empirical results may serve in the future for validating the theoretical models. Indeed such a validation would be very welcome and useful to use the models for comparison with observations in environments in which the number of Cepheids is not high enough to empirically determine the IS edges. For example, such observationally calibrated models could serve to identify Cepheids in SN host galaxies.
According to the comparison shown in this paper, currently, models by \citet{Anderson2016} with either high rotation rates or a mixture of moderate and high rotation rates, depending on the pulsation period, provide the best agreement with the obtained empirical instability strip.

As the OGLE catalog of Cepheids in the LMC has very high completeness we do not expect practically any improvement in the statistics in the future maybe except for the faintest Cepheids. Nevertheless, in any future study, an improvement in the reddening determination and, to a lower extent, photometric precision (mostly for faint stars) would help to obtain more accurate IS edges for the LMC galaxy.
At the moment, however, the best way to get more insight into the physics of the instability strip is to study Cepheids in other galaxies. Not only does it provide a different and independent sample of stars but it also lets us check on the effect of metallicity, which has a significant influence on the stellar evolution and pulsation.
For that reason, our next step will be to perform a similar analysis for the SMC galaxy and compare the results.

\begin{acknowledgements}
We thank the anonymous referee for the constructive comments and suggestions. We acknowledge support from the Polish National Science Center grant SONATA BIS 2020/38/E/ST9/00486. RS is supported by the National Science Center, Poland, SONATA BIS project 2018/30/E/ST9/00598. This research has made use of NASA's Astrophysics Data System Service.
\end{acknowledgements}

%
%
\bibliographystyle{aa.bst}
\bibliography{main}

\end{document}